\documentclass[aps,prb,superscriptaddress,twocolumn]{revtex4}

\usepackage{amsmath,amssymb}
\usepackage{graphicx}
\usepackage{feynmf}
\usepackage{color}
\usepackage{dsfont}

\newcommand{\rs}{\scriptscriptstyle\rm(s)}
\newcommand{\rd}{\scriptscriptstyle\rm (d)}
\newcommand{\ri}{\scriptscriptstyle\rm (i)}
\newcommand{\s}{\scriptscriptstyle}
\newcommand{\sr}{\scriptscriptstyle\rm}
\begin{document}

\title{Heat generation due to the Anderson catastrophe in mesoscopic devices}

\date{\today}

\author{A.V.\ Lebedev}
\affiliation{Moscow Institute of Physics and Technology, Institutskii per.\ 9, Dolgoprudny,
141700, Moscow District, Russia}
\affiliation{Dukhov Research Institute of Automatics (VNIIA), Moscow,
Russia, 127055}

\author{V.M.\ Vinokur}
\affiliation{Materials Science Division, Argonne National Laboratory,
		9700 S. Cass Avenue, Argonne, IL 60439, USA}

\begin{abstract}
Anderson's orthogonality catastrophe (AOC) theorem establishes that the ground state of the many-body fermion system is asymptotically orthogonal to the ground state of the same system perturbed by a scattering potential, so that the overlap between the original and new ground states decays to zero with the system size.  We adopt the AOC for a description of heat production in a complementary metal-oxide-semiconductor (CMOS) transistor. We find that the heat released in the transistor comprises two distinct components, contribution from the dissipation accompanying electron transmission under the applied voltage and purely quantum-mechanical AOC part due to the change in scattering matrix for electrons upon switching between high and low conductance regimes. We calculate the AOC-induced heat production, which we call switching heat.
\end{abstract}

\maketitle

In 1967, Anderson\,\cite{Anderson1967} discovered that in the limit $N\to\infty$, the ground state of a free gas of $N$ non-interacting fermions is orthogonal to the ground state of the same system but perturbed by an external potential. The asymptotic overlap of the two ground states behaves as $N^{-\gamma}$, where the $\gamma>0$ constant is determined by the parameters of the scattering potential. As has been already pointed out by Anderson, such a response of an electron gas to the external potential leads to the creation of electron-hole excitations and thus heating. This phenomenon, which is realized now to be an inherent phenomenon in general many-body systems subject to an abrupt perturbation, is termed Anderson's orthogonality catastrophe (AOC).  The AOC theorem originally proved for the system of a noninteracting electron gas with a separable local potential, has been extended to more general cases with a local potential and electron interactions and applies to diverse problems in physics. Several AOC's most interesting and dramatic consequences occur in the realm of nonequilibrium dynamics and quantum work statistics\,\cite{Binder2018}. This aspect of AOC has been successfully adopted to describe the Kondo effect and the edge singularity of x-ray absorption and emission in metals\,\cite{Binder2018} and vortex dynamic dissipation\,\cite{Finn}.

In nanodevices, such as  a single electron transistor\,\cite{SET}, as well as in nanoscale digital semiconductor devices, the change in the scattering potential leads to the readjustment of electron scattering states in the leads and, thus, the AOC phenomenon has to manifest. A major challenge hindering the progress of semiconductor technology is the enhanced energy consumption of integrated semiconductor-based electronic devices\,\cite{pop2010,solomon2010,esseni2017}.  An exemplary system exposing the problem is a single central processing unit (CPU), which at the GHz operating frequency suffers about 100-W/cm$^2$ waste heat density generation. Here we apply the AOC concept to calculate the heat production accompanying the electron transmission in an elementary building block of a modern CPU, the metal-oxide-semiconductor (CMOS) transistor. We note, however, that the obtained results is much more general and can be applied to an arbitrary mesoscopic device with a varying scattering potential.

\begin{figure}[ht]
  \centering
  \includegraphics[width=8cm]{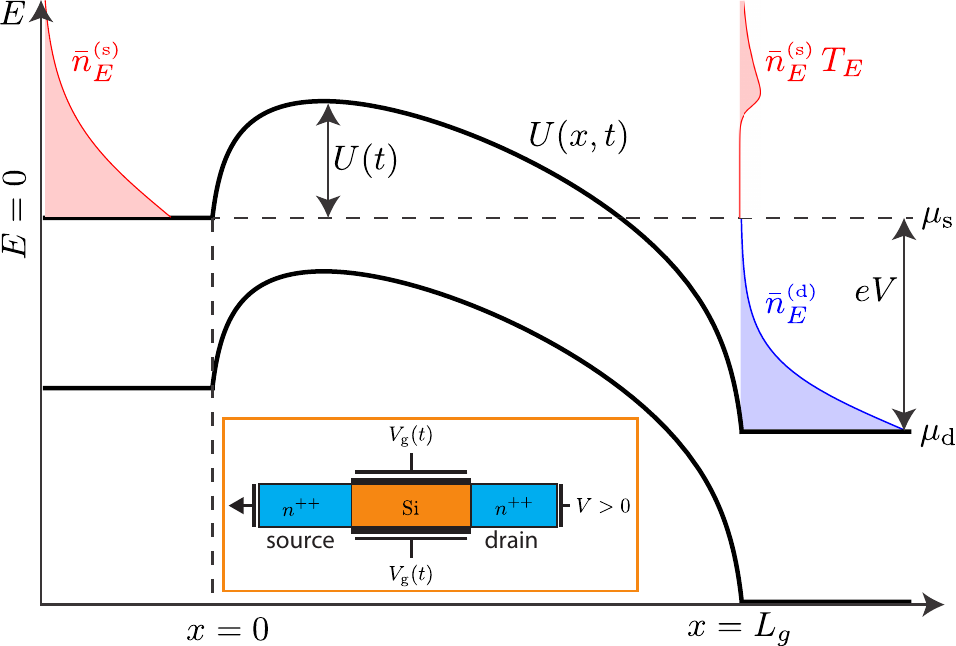}
  \caption{ Main panel: The energy band structure of the n$^{++}$-Si-n$^{++}$ metal oxide field effect transistor. The hot electrons in the source (at $x<0$) with the energy distribution $\bar{n}^{\rs}_E$ (red) propagate towards the gate region, $x\in[0,L_{\mathrm g}]$, where the electrostatic potential barrier $U(x,t)$ with the barrier height $U(t)$ is formed. The source electrons are transmitted to the drain contact $x>L_{\mathrm g}$ with the probability $T_E$ and relax to the thermal equilibrium state with the chemical potential shifted by the supply voltage $V>0$. The drain electrons with the energy distribution $\bar{n}^{\rd}_E$ (blue) do not nave enough energy to tunnel into the source and are mostly back-reflected to the drain. The inset: A general setup.
   }\label{fig:setup}
\end{figure}

The CMOS\,\cite{szhe,taur} is a three-terminal semiconductor device, see Fig.\,\ref{fig:setup} with the resistance between the source and the drain being externally controlled via the voltage $V_{\mathrm g}(t)$ applied to the third terminal, the gate. In the fully open, `ON,' regime under the finite voltage $V$ applied to the drain, an electric current $I_\mathrm{on}$ flows across the system, and the corresponding  Joule heat power $I_\mathrm{on} V$ is released. In the high resistance, the `OFF,' regime the leakage power $I_\mathrm{off} V$ is generated. The nature of this Joule heating is purely classical. The excess high energy electrons incoming from the source to the voltage biased drain dissipate its energy due to electron-phonon collision processes. In large scale transistors with the size beyond $100$\,nm, its electrical and heat characteristics can be found within a classical drift-diffusion model of electronic transport\,\cite{szhe,taur}. In smaller devices, one has to take into account the electron-phonon scattering adopting the nonequilibrium Green's function formalism\,\cite{lake1992}. From the practical viewpoint, a seemingly obvious way to minimize the heat production is to decrease the operating voltage $V$. Indeed, based on the Landauer erasure principle\,\cite{landauer1961} it was shown\,\cite{meindl} that a minimal supply voltage needed to switch an electronic inverter is $V = 2\ln(2) k_{\mathrm{\scriptscriptstyle B}}T/|e|$. However the resulting process appears too slow and cannot be used in practice for fast digital applications.

It is miniaturization of transistors down to $10$\,nm scales that brings to the stage quantum phenomena. A geometrical quantization of electronic levels becomes visible\,\cite{stern1967,stern1972}, and the transport is mediated by discrete modes. At the $10$\,nm scale, the size of the device becomes smaller than the electron-phonon scattering length, and charge carriers ballistically travel from source to drain\,\cite{likharev}. Hence, the Joule heating formula ${\cal Q} = j(x) {\cal E}(x)$ for the local heat power density, where $j(x)$ is current density and ${\cal E}(x)$ is the local electric field, does not straightforwardly apply.  But then the heat dissipation occurs mostly inside the device terminals where no electric field is present, enables us to use, instead, the scattering matrix approach\,\cite{blanter} for the nanometer size transistor devices\,\cite{natori,lundstrom1997,lundstrom2002}. To that end, we note, as a first step, that the electronic coherence comes into play so that transmission electrons belong in a many-body quantum superposition state. Utilizing the scattering matrix approach, we derive an expression for the total heat release in a one-dimensional mesoscopic conductor, assuming the ballistic propagation inside the scattering region and relaxing to the thermal equilibrium via the electron-phonon interaction within the device leads. We find that the released heat comprises two distinct contributions: (i)\,The heat from the electron transmission under the applied voltage and (ii)\,the purely quantum-mechanical contribution due to the change in the scattering matrix of the device upon switching between high and low conductance regimes.  The former contribution represents the standard Joule heating associated with electronic transport.  The latter, which we refer to as to switching heating, originates from the AOC phenomenon, which, in its extreme manifestation, is a sudden change in the scattering matrix, so that the wave functions of an underlying Fermi sea electron have to readjust to the new potential.

The AOC results in a creation of electron-hole pairs in the system thus taking the device terminals out of the local equilibrium state. The switching heat is then generated during the relaxation of these electron-hole pairs back to the equilibrium state. Importantly, this contribution is always present even with no supply voltage applied, $V=0$, and is inversely proportional to the switching time of the scattering matrix $\propto \hbar/\tau_\mathrm{sw}$. Therefore, this contribution might be a new limiting factor for the emerging generation of fast digital electronic transistors.

We consider a non-stationary process where the time-dependent scattering matrix $\hat{S}(t)$ of the electrostatic barrier of the device switches from the low conducting `OFF'- regime $\hat{S}(t) = \hat{S}^{\scriptscriptstyle (0)}$ to the high conducting `ON' state $\hat{S}(t) = \hat{S}^{\scriptscriptstyle (1)}$ within a finite time window $t\in[-\Delta t/2, \Delta t/2]$. Here we assume an adiabatic switching where the tunneling time $\propto \hbar \partial_E T_E$ ($T_E$ is a transmission probability between the contacts) and a ballistic traveling time through the scattering region are both much shorter than the switching time $\tau_\mathrm{sw}$. We also assume that the source and drain terminals are in a thermal equilibrium at the temperature $T$ with Fermi potentials $\mu_\mathrm{s}$ and $\mu_\mathrm{d} = \mu_\mathrm{s} - |e|V$, respectively, where $V>0$ is a supply voltage applied to the device, see Fig.\ref{fig:setup}. The released heat is generated via equilibration of the excess carriers redistributed between the terminals. In case the terminals rest at constant temperature the heat is given by the Clausius law:
\begin{equation}
       Q = k_{\scriptscriptstyle{\mathrm B}} T\, \bigl[ \delta S_\mathrm{s} + \delta S_\mathrm{d} \bigr],
       \label{eq:Clausius}
\end{equation}
where $\delta S_\mathrm{s,d}$ is the excess entropy generated inside the source and drain reservoirs. For a non-interacting coherent Fermi gas, its entropy coincides with von Nuemann entropy and can be expressed through a single-electron density matrix $S = -\mbox{Tr}\{ \hat\rho \ln(\hat\rho)\}$. In practice, electrons in the terminals experience strong dephasing due to interaction with low energy phonon modes and therefore the electron density matrix gets diagonal in the energy representation, $\bigl[\hat\rho\bigr]_{E,E^\prime} \to \delta(E-E^\prime) \bigl[\hat\rho\bigr]_{E,E}$. Hence the excess entropy can be expressed solely through the electron occupation numbers, $n_{\epsilon_k}$, as
\begin{equation}
      S_\mathrm{i} = - \int \frac{dk L}{2\pi} \Bigl[ n_{\epsilon_k}^{\ri} \ln n_{\epsilon_k}^{\ri} + (1-n_{\epsilon_k}^{\ri}) \ln(1-n_{\epsilon_k}^{\ri}) \Bigr],
      \label{eq:entropy}
\end{equation}
where $L\to\infty$ is the length (volume) of the gas, i$\in\mathrm{s,d}$.  In the `ON' regime, the scattering matrix of the device is more transparent and redistributed excess electrons modify the electron occupation numbers $n_{\epsilon}^{\ri} \to \bar{n}_\epsilon^{\ri} + \delta n_\epsilon^{\ri}$, where $\bar{n}_\epsilon^{\ri}$ is an equilibrium electron occupation number, $\bar{n}_\epsilon^{\ri} = \bigl(1+e^{(\epsilon-\mu_\mathrm{i})/k_BT}\bigr)^{-1}$. Assuming $|\delta n_\epsilon^{\ri}| \ll \bar{n}_\epsilon^{\ri}$, one can expand the entropy, see Eq. (\ref{eq:entropy}), to the first order in $\delta n_\epsilon^{\ri}$ and get the expression for the entropy change,
\begin{eqnarray}
      \delta S_\mathrm{i} = - \int \frac{dk L}{2\pi} \, \delta n_{\epsilon_k}^{\ri} \ln\Bigl[ \frac{\bar{n}_{\epsilon_k}^{\ri}}{1-\bar{n}_{\epsilon_k}^{\ri}} \Bigr].
      \label{eq:dS}
\end{eqnarray}
Substituting the equilibrium occupation numbers into Eq.(\ref{eq:dS}) the Clausius law, Eq.(\ref{eq:Clausius}) gives the released heat
\begin{equation}
      Q = \sum\limits_{\mathrm{i}\in\{\mathrm{s,d}\}} \int \frac{dk L}{2\pi}\,  (\epsilon_k - \mu_\mathrm{i})\,\delta n_{\epsilon_k}^{\ri}.
      \label{eq:Q}
\end{equation}

To find the variation in the occupation numbers $\delta n_\epsilon^{\ri}$, we make use of the non-stationary electron scattering states in the Wentzel-Kramers-Brillouin (WKB) approximation for the source $\varphi_E^{\rs}(x,t)$ and drain $\varphi_E^{\rd}(x,t)$ reservoirs, see Appendix \ref{A1} for details. At $t\ll-\Delta t/2$ and $t\gg\Delta t/2$, the non-stationary scattering states coincide with the stationary scattering states $\bar{\varphi}^{\rs}(x,t)$ and $\bar{\varphi}^{\rd}(x,t)$ corresponding to the 'OFF' regime of the device. Both nonstationary and stationary states represent a complete set of the orthogonal states and, therefore, the electronic field operator in the device terminals can be equivalently represented through each of these sets as
\begin{eqnarray}
      \hat\Psi(x,t) &=& \sum_E \varphi^{\rs}_E(x,t) \hat{\bar a}_E + \varphi^{\rd}_E(x,t) \hat{\bar b}_E,
      \label{eq:Psi}
      \\
      \hat\Psi(x,t) &=& \sum_E \bar{\varphi}^{\rs}_E(x,t) \hat{a}_E(t) + \bar{\varphi}^{\rd}_E(x,t) \hat{b}_E(t),
      \label{eq:Psi0}
\end{eqnarray}
where $\hat{\bar{a}}_E$ and $\hat{\bar{b}}_E$ denote stationary electron annihilation operators at energy $E$ incoming from the source and drain reservoirs with $\langle \hat{\bar a}_E^\dagger \hat{\bar a}_E\rangle = \bar{n}^{\rs}_E$ and $\langle \hat{\bar b}_E^\dagger \hat{\bar b}_E\rangle = \bar{n}^{\rd}_E$. The new non-stationary operators $\hat{a}_E(t)$ and $\hat{b}_E(t)$ give the non-equilibrium occupation numbers $n_E^{\rs} = \langle \hat{a}_E^\dagger \hat{a}_E\rangle$ and $n_E^{\rd} = \langle \hat{b}_E^\dagger \hat{b}_E\rangle$ which capture the presence of electron-hole excitations in the terminals due to the change in the scattering potential. Making use of the orthogonality of the scattering states, one gets
\begin{equation}
      \left[\begin{array}{c} \hat{a}_E(t)\\ \hat{b}_E(t) \end{array}\right] = \sum_{E^\prime}
      \!\!\left[\begin{array}{cc}
      (\varphi^{\rs}_{E^\prime} \cdot \bar\varphi^{\rs}_E)& (\varphi^{\rd}_{E^\prime} \cdot \bar\varphi^{\rs}_E)\\
      (\varphi^{\rs}_{E^\prime} \cdot \bar\varphi^{\rd}_E)& (\varphi^{\rd}_{E^\prime} \cdot \bar\varphi^{\rd}_E)
      \end{array}\right]\!\!
      \left[ \begin{array}{c}
      \hat{\bar a}_{E^\prime}\\ \hat{\bar b}_{E^\prime}
      \end{array}\right],
      \label{eq:ab_transform}
\end{equation}
where $(\varphi^{\ri}_{E^\prime} \cdot \bar\varphi_E^{\scriptscriptstyle\rm (j)}) \equiv \int dx \varphi^{\ri}_{E^\prime}(x,t) [\bar\varphi^{\scriptscriptstyle\rm (j)}_E(x,t)]^*$ with $\mathrm{i,j}\in\{\mathrm{s,d}\}$ are the scattering states overlaps.

The variation of the electron occupation numbers are derived in Appendix \ref{A1}, see Eqs.(\ref{eq:dnA}) and (\ref{eq:dnB}). In order to understand the result, let us consider a situation where switching time of the potential barrier, $\tau_\mathrm{sw}$ is much smaller than the duration of the `ON' regime: $\tau_\mathrm{sw}$$\ll $$\Delta t$. Then for $\tau$$\in$$[-\Delta t/2,\Delta t/2]$ one assumes scattering amplitudes to be constant in time, and the variation of the occupation numbers can be decomposed into the equilibrium and non-equilibrium parts: $\delta n^{\s (\mathrm{s,d})}_E = \bigl[\delta n^{\s(\mathrm{s,d})}_E\bigr]_\mathrm{eq} + \bigr[ \delta n^{\s(\mathrm{s,d})}_E \bigr]_\mathrm{neq}$. The equilibrium part is finite when no supply voltage is applied to the device, $V=0$,
\begin{eqnarray}
      &&\bigl[\delta n_E^{\rs} \bigr]_\mathrm{eq} \!=\! \frac{v_E^{\rs}}{L} \!\int\!\! \frac{dE^\prime}{2\pi\hbar} \int d\tau d\tau^\prime \, e^{i(E^\prime-E)(\tau-\tau^\prime)/\hbar}
      \label{eq:eq}\\
      &&\quad\times  \bar{n}_{E^\prime}^{\rs} \bigl[ r_{E^\prime}^*(\tau)r_{E^\prime}(\tau^\prime) + t_{E^\prime}^*(\tau) t_{E^\prime}(\tau^\prime)-1\bigr],
      \nonumber
\end{eqnarray}
and similarly for the drain reservoir with $r_{E^\prime}\to \bar{r}_{E^\prime}$.

We argue that the density variation $\bigl[\delta n_E^{\rs} \bigr]_\mathrm{eq}$ is indeed associated with the AOC phenomenon as far as it describes a redistribution of the electron density due to shaking up of the Fermi sea by the change in the scattering potential. First, we note, that the equilibrium density variation does not bring additional electron charges to the contacts. Integrating  $[\delta n_E^{\rs}]_\mathrm{eq}$ over the energy, one gets the net equilibrium charge variation in the source reservoir, $\bigl[\delta N^{\rs}\bigr]_\mathrm{eq} = \sum_E \bigl[ \delta n_E^{\rs} \bigr]_\mathrm{eq} \equiv \int dE\, L \bigl[ \delta n_E^{\rs} \bigr]_\mathrm{eq}/(2\pi\hbar v_E^{\rs})$. The $dE$ integration gives the $\delta$-function $\delta(\tau-\tau^\prime)$ under the double time integral in Eq.(\ref{eq:eq}) which then vanishes due to normalization of the scattering amplitudes: $|t_E(\tau)|^2 + |r_E(\tau)|^2=1$. Thus, $\bigl[ \delta N^{\rs} \bigr]_\mathrm{eq}$$=$$0$ and Eq.(\ref{eq:eq}) describe the density variation due to electron-hole excitations created in the source terminal. Second, the main contribution to the double time integral in Eq.(\ref{eq:eq}) comes from the left $\tau \lesssim -\Delta t/2$, $\tau^\prime \gtrsim -\Delta t/2$  and right $\tau \gtrsim \Delta t/2$, $\tau^\prime \lesssim \Delta t/2$ switching edges where the scattering matrix changes and does not depend on $\Delta t$. Therefore, as we will see below, the corresponding heat contribution is proportional to a change rate of the scattering matrix what confirms its AOC origin.

The non-equilibrium contributions at $\Delta t  \gg \tau_\mathrm{sw}$ are given by
\begin{equation}
      \bigl[ \delta n_E^{\s\rm(s,d)} \bigr]_\mathrm{neq} \!\!=\! \mp\frac{v_E^{\s\rm (s,d)} \Delta t}{L}\, \bigl[ \bar{n}_E^{\rs} \!-\! \bar{n}_E^{\rd} \bigr]  \bigl| t_E^{\s(1)} \bar{r}_E^{*\s(0)} \!+\! r_E^{\s(1)} t_E^{*\s(0)}\bigr|^2,
      \label{eq:neq}
\end{equation}
and are proportional to the time duration $\Delta t$ of the `ON' regime and vanish at $V$$=$$0$. Thus, $\bigl[ \delta n_E^{\s\rm(s,d)}\bigr]_\mathrm{neq}$ describes the transport contribution to the electron density variation related to electron flow between source and drain reservoirs. Substituting Eq.\,(\ref{eq:neq}) into Eq.\,(\ref{eq:Q}) one gets an associated heat released in the terminals:
\begin{equation}
      Q_\mathrm{tr}^{\s\rm(s,d)} \!= \mp\!\! \int\!\! \frac{dE \Delta t}{2\pi\hbar}\, (E-\mu_\mathrm{s,d}) \bigl[ \bar{n}_E^{\rs} \!-\! \bar{n}_E^{\rd} \bigr]  \bigl| t_E^{\s(1)} \bar{r}_E^{*\s(0)} \!+\! r_E^{\s(1)} t_E^{*\s(0)}\bigr|^2.
\end{equation}
One notes that at $\mu_\mathrm{s} > \mu_\mathrm{d}$ i.e. at $V>0$, the released heat in the source can be negative if the transmission factor $\bigl| t_E^{\s(1)} \bar{r}_E^{*\s(0)} \!+\! r_E^{\s(1)} t_E^{*\s(0)}\bigr|^2 \to 1$ at $E>\mu_\mathrm{s}$ and vanishes at $E<\mu_\mathrm{s}$. This cooling effect can be explained by depletion of the hot carriers in the source reservoir due to their transfer into the drain, see Fig.\ref{fig:setup}. These hot electrons transmitted to the drain heat up the drain terminal where the released heat gets positive.
Summing up the source and drain heat together one gets the net transport heat released in the system as
\begin{equation}
      Q_\mathrm{tr} = |e|V \!\!\int\!\! \frac{dE \Delta t}{2\pi\hbar}\, \bigl( \bar{n}_E^{\rs} \!-\! \bar{n}_E^{\rd} \bigr) \bigl| t_E^{\s(1)} \bar{r}_E^{*\s(0)} \!+\! r_E^{\s(1)} t_E^{*\s(0)}\bigr|^2,
      \label{eq:Qsd}
\end{equation}
which is always positive. Moreover, integrating the variation of the source electron occupation number, see Eq.(\ref{eq:neq}), over energy, one gets the total charge $q = e\sum_E \delta n_E^{\rs}$ transmitted from the source to drain, and restores  the Joule heating law $Q^\mathrm{tr} = |q| V_\mathrm{sd}$.

The above results, see Eqs.(\ref{eq:neq}) and (\ref{eq:Qsd}), are derived for a longtime opening regime $\Delta t \gg \tau_\mathrm{sw}$.  Let us consider a more general case of an arbitrary time dependence of the scattering coefficients. Here we focus on a practical situation of the highly doped semiconductor, where the Fermi levels $\mu_\mathrm{s,d}$ coincide with the bottom of the conduction bands $E_\mathrm{c}^{\scriptscriptstyle\rm (s,d)}$. The released heat can be decomposed into a sum of the switching component $Q_\mathrm{sw}$ which is finite at $V=0$ and the transport component $Q_\mathrm{tr}$ which vanishes at $V=0$: $Q=Q_\mathrm{sw}+ Q_\mathrm{tr}$,
\begin{widetext}
\begin{eqnarray}
      &&Q_\mathrm{sw} = \!\!\int\!\! \frac{dE}{2\pi} \!\!\int\!\! \frac{dt dt^\prime}{2\pi} \Bigl\{ \bar{n}_E^{\rs} \frac{1\!-\!r_E^*(t) r_E(t^\prime)\!-\!t_E^{*}(t) t_E(t^\prime)}{(t-t^\prime-i\delta)^2} e^{i\frac{E}{\hbar}(t\!-\!t^\prime)} \!\!+ \!\bar{n}_E^{\rd}\frac{1\!-\!\bar{r}_E^*(t) \bar{r}_E(t^\prime)\!-\!t_E^{*}(t) t_E(t^\prime)}{(t-t^\prime-i\delta)^2} e^{i\frac{(E-eV)}\hbar(t\!-\!t^\prime)}\Bigr\},
      \label{eq:Qsw}
      \\
      &&Q_\mathrm{tr}= \!\!\int\!\! \frac{dE}{2\pi} \!\!\int\!\! \frac{dt dt^\prime}{2\pi} \frac{1-e^{-ieV(t-t^\prime)/\hbar}}{(t-t^\prime-i\delta)^2} e^{iE(t-t^\prime)/\hbar} \Bigl( \bar{n}_E^{\rs} f_E^{*\rs}(t) f_E^{\rs}(t^\prime) - \bar{n}_E^{\rd} f_E^{*\rd}(t) f_E^{\rd}(t^\prime)\Bigr),
      \label{eq:Qtr}
\end{eqnarray}
\end{widetext}
where $f_E^{\rs}(t) = t_E(t)\bar{r}_E^{*\s (0)} +r_E(t) t_E^{*\s (0)}$, $f_E^{\rd}(t) = t_E(t) r_E^{*\s (0)}+\bar{r}_E(t) t_E^{*\s(0)}$ and $\delta >0$ is the regularization constant.

The released switching heat, see Eq.(\ref{eq:Qsw}), is associated with the time-dependence of the scattering coefficients at the switching edges $t=\pm \Delta t/2$. At large $V \gg k_BT/e$, the scattering coefficients typically do not sufficiently change at $E \sim \mu_\mathrm{d}$ and the main contribution to the switching heat is given by the term $\propto \bar{n}_E^{\rs}$ in Eq. (\ref{eq:Qsw}). Making use of the normalization constraint $r_E^*(t)r_E(t) + t_E^*(t) t_E(t) = 1$, one can transform the switching heat contribution into a regular form
\begin{eqnarray}
      &&Q_\mathrm{sw} = \frac12 \int dE \,\bar{n}_E^{\rs} \int\frac{dtdt^\prime}{(2\pi)^2} \, \Bigl\{ \frac{r_E(t)-r_E(t^\prime)}{t-t^\prime}
      \label{eq:Qsw2}\\
      &&\quad \times \frac{r^*_E(t) e^{iE(t\!-\!t^\prime)/\hbar}\!- r^*_E(t^\prime) e^{-iE(t\!-\!t^\prime)/\hbar}}{t-t^\prime}
      \nonumber\\
      &&+ \frac{\bigl[t_E(t)\!-\!t_E(t^\prime)\bigr]\! \bigl[t^*_E(t) e^{iE(t\!-\!t^\prime)/\hbar}\!\!-\! t^*_E(t^\prime) e^{-iE(t\!-\!t^\prime)/\hbar}\bigr]}{(t-t^\prime)^2}\Bigr\}.
      \nonumber
\end{eqnarray}

Until now, we have implicitly assumed that the device is fully coherent during its operation and only at $t\gg \Delta t/2$ the off-diagonal elements of the electron density matrix $\rho_{E,E^\prime}$ vanish due to the weak decoherence process. In reality, transistors operate at high temperatures $T\simeq 300$\,K in a strong decoherence regime. Therefore, the time difference $t-t^\prime$ in the double time integrals in Eqs.\,(\ref{eq:Qsw}) and (\ref{eq:Qtr}) has to be restricted by the decoherence time $\tau_\varphi$ via a damping factor $e^{-|t-t^\prime|/\tau_\varphi}$. Let us consider the adiabatic situation, where the scattering amplitudes are the slowly varying functions at the $\tau_\varphi$ time scale
\begin{eqnarray}
      &&Q_\mathrm{sw} \approx -\!\!\int\!\! \frac{dE}{\pi^2}\, \bar{n}_E^{\rs} \arctan\Bigl( \frac{E\tau_\varphi}{\hbar}\Bigr)\!\!\int\! dt \bigl( R_E \dot\varphi_E \!+\! T_E \dot\theta_E\bigr)
      \nonumber\\
      &&+ \!\int\!\! \frac{dE}{(2\pi)^2}\, \bar{n}_E^{\rs}\,\frac{\hbar^2/\tau_\varphi}{(\hbar/\tau_\varphi)^2 +E^2} \int\! dt \Bigl( |\dot{r}_E|^2 + |\dot{t}_E|^2 \Bigr),
      \label{eq:Qsw3}
\end{eqnarray}
where $\varphi_E$ and $\theta_E$ are scattering phases defined through $r_E=\sqrt{R_E} e^{i\varphi_E}$ and $t_E = \sqrt{T_E} e^{i\theta_E}$ and
$R_E(t) = |r_E(t)|^2$ and $T_E(t) = |t_E(t)|^2$ are reflection and transmission probabilities for source electrons. The time dependencies of scattering amplitudes are controlled via the gate voltage $V_{\mathrm g}(t)$. Therefore, the first term in Eq.(\ref{eq:Qsw3}) can be transformed into the form, $\int dt \bigl(R_E  \dot\varphi_E + T_E \dot\theta_E\bigr) = \int \bigl( R_E(V_{\mathrm g}) d \varphi_E(V_{\mathrm g}) + T_E(V_{\mathrm g}) d \theta_E(V_{\mathrm g})\bigr)$, which vanishes for $V_{\mathrm g}(t<-\Delta t/2) = V_{\mathrm g}(t>\Delta t/2)$. Thus, the switching heat is given by the second positive term in Eq.(\ref{eq:Qsw3}).

To gain insight into the quantitative dependence of the switching heat upon scattering potential parameters, we consider a model with the rectangular shape  of the electrostatic potential barrier: $U(x,t) = U(t)$ for $x\in[0,L_g]$ and $U(x,t)=0$ outside the gate region. The height of the barrier $U(t)$ is time dependent and controlled via the gate voltage $V_g(t)$. The corresponding scattering amplitudes for $E>U$ assume the form
\begin{eqnarray}
      &&r_E\!=\!\frac{(k_E^2-\tilde{k}_E^2)(e^{i\tilde{k}_EL_g}-e^{-i\tilde{k}_EL_g})}{(\tilde{k}_E\!-\!k_E)^2 e^{i\tilde{k}_EL_g}\!-\!(\tilde{k}_E\!+\!k_E)^2 e^{-i\tilde{k}_E L_g}},\>\>\>
      \\
      &&t_E\!=\!\frac{-4k_{E}\tilde{k}_Ee^{-ik_E L_g}}{(\tilde{k}_E\!-\!k_E)^2 e^{i\tilde{k}_EL_g}\!-\!(\tilde{k}_E\!+\!k_E)^2 e^{-i\tilde{k}_E L_g}},\>\>\>
\end{eqnarray}
where $k_E = \sqrt{2mE}/\hbar$ and $\tilde{k}_E = \sqrt{2m(E-U)}/\hbar$. The underbarrier scattering amplitudes are given by the similar expressions after the substitution $\tilde{k}_E \to -i\kappa_E$ with $\kappa_E = \sqrt{2m(U-E)}/\hbar$. The right hand side scattering amplitude can be obtained from the unitarity constraint $\bar{r}_E=-r_E^* t_E/t_E^*$.

\begin{figure}[ht]
  \centering
  \includegraphics[width=6cm]{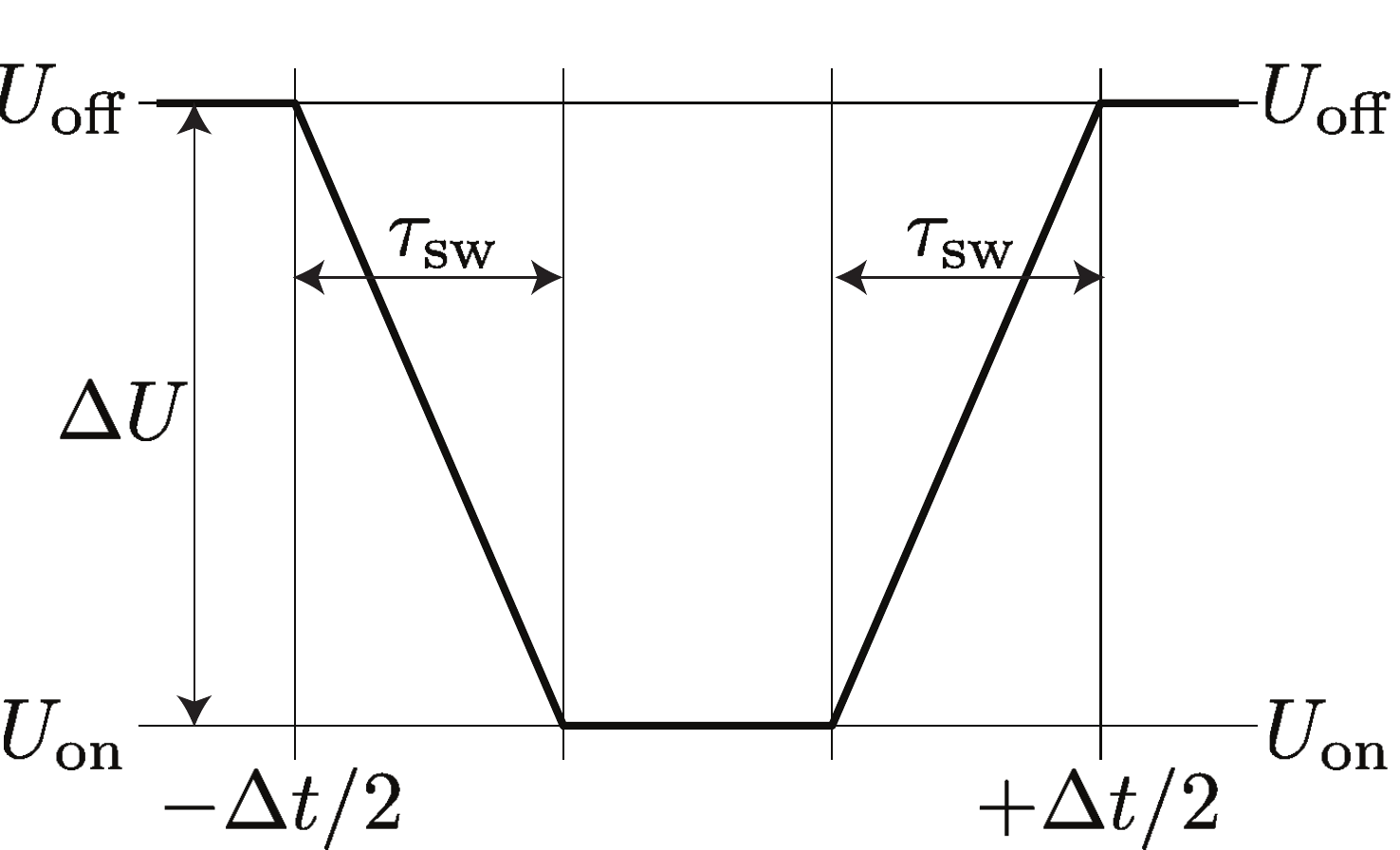}
  \caption{Time-dependent profile of the potential barrier height.}\label{fig:gate}
\end{figure}

We consider a non-biased situation $eV \to 0$ and assume that the potential barrier height $U(t)$ drops linearly from the value $U_\mathrm{off}$ locking the contact  to the opening value $U_\mathrm{on}< U_\mathrm{off}$ during the switching time $\tau_\mathrm{sw}$ at the left edge $t=- \Delta t/2$ of the gate voltage pulse. Similarly, potential restores back linearly to $U(t)=U_\mathrm{off}$ at the right edge $t=\Delta t/2$, see Fig.\ref{fig:gate}.
Given this linear barrier height time-dependence, the switching heat Eq.(\ref{eq:Qsw3}) becomes
\begin{equation}
      Q_\mathrm{sw}= \frac\hbar{\tau_\mathrm{sw}}\, \Delta u \, F(\eta,u_\mathrm{on},\theta),
\end{equation}
where $\Delta u = (U_\mathrm{off}-U_\mathrm{on})/k_{\sr B}T$, and $F$ is a dimensionless factor,
\begin{equation}
      F(\eta,u_\mathrm{on},\theta) = \!\!\! \int\limits_0^{\infty}\!\! \frac{d\epsilon}{\pi^2} \frac{\eta\,n_F(\epsilon)}{\eta^2\!+\!\epsilon^2} \!\!\int\limits_{u_\mathrm{on}}^{u_\mathrm{off}}\!\! du
      \Biggl[ \Bigl|\frac{dr(\epsilon,\theta)}{du}\Bigr|^2 \!\!+\! \Bigl|\frac{dt(\epsilon,\theta)}{du}\Bigr|^2\Biggr].
      \label{eq:heat_rec}
\end{equation}
Here, $\epsilon$ and $u$ are electron energy and potential barrier height measured in $k_{\sr B}T$ units, $\eta = \tau_{\sr T}/\tau_\varphi$ is a ratio of the temperature time $\tau_T=\hbar/k_{\sr B}T$ and the dephasing time, $\theta = \sqrt{2mk_{\sr B} T}L_{\mathrm g}/\hbar$ is the phase accumulated by a charge carrier with the mass $m$ and thermal energy $k_{\sr B}T$ during its propagation through the gate region, $n_{\sr F}(\epsilon) = 1/(1+e^\epsilon)$.

\begin{figure}[ht]
  \centering
  \includegraphics[width=8.5cm]{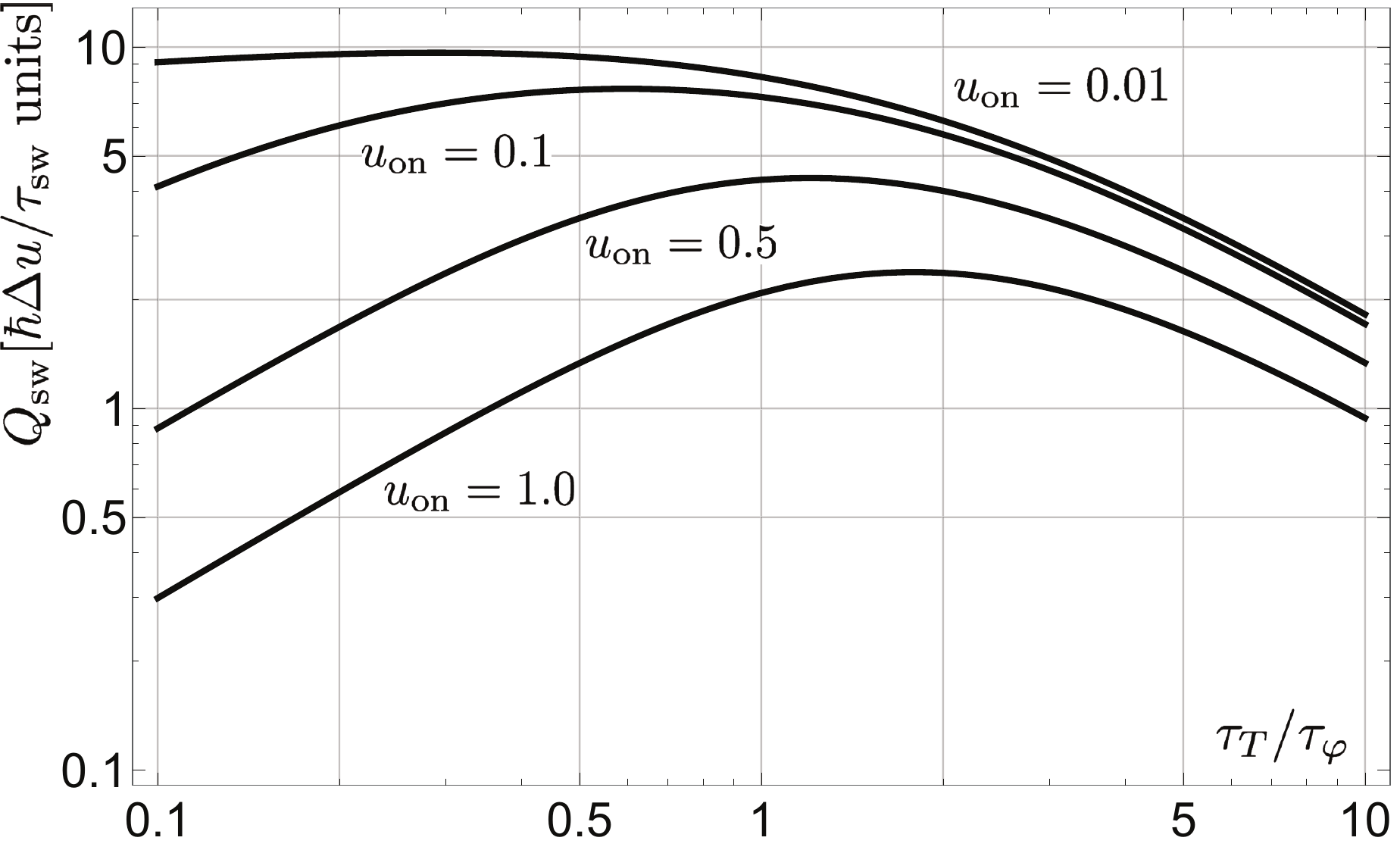}
  \caption{The released switching heat in $\hbar \Delta u/\tau_\mathrm{sw}$ energy units (vertical axis, $\log$ scale)  versus the dephasing rate $\tau_{\scriptscriptstyle{\mathrm T}}/\tau_\varphi$ (horizontal axis, $\log$ scale) for four different potential barrier heights $u_\mathrm{on}$.}\label{fig:q_vs_eta1}
\end{figure}

As follows from Eq.\,(\ref{eq:heat_rec}), the value of the switching heat is set by the Heisenberg energy scale $\hbar/\tau_\mathrm{sw}$ associated with the switching time of the potential barrier. This indicates a quantum character of this heat contribution. Moreover, the switching heat depends explicitly  on the decoherence time $\tau_\varphi$ through the parameter $\eta$, see Fig.\,\ref{fig:q_vs_eta1}. As far as the coherence of the system grows $\tau_\varphi\to \infty$ ( i.e. $\eta\to 0$) the switching heat decreases and formally vanishes at $\eta = 0$. Note, however, that Eq.\,(\ref{eq:Qsw3}) is valid only for $\tau_\mathrm{sw} \gg \tau_\varphi$. In the opposite limit $\tau_\varphi \gg \tau_\mathrm{sw}$ one approaches a non-adiabatic regime where Eq.\,(\ref{eq:Qsw3}) is not valid any more, and one has to use Eq.\,(\ref{eq:Qsw2}) instead. For fast digital applications, the height of the potential barrier in the `ON' regime has to be small in order to charge the drain capacitor quickly. As the potential barrier height decreases, $u_\mathrm{on}$$\to$$0$, the released heat increases, see Fig.\,\ref{fig:q_vs_eta1}, as more and more electrons get involved into the AOC near the bottom of the conduction band.

\begin{figure}[ht]
  \centering
  \includegraphics[width=8.5cm]{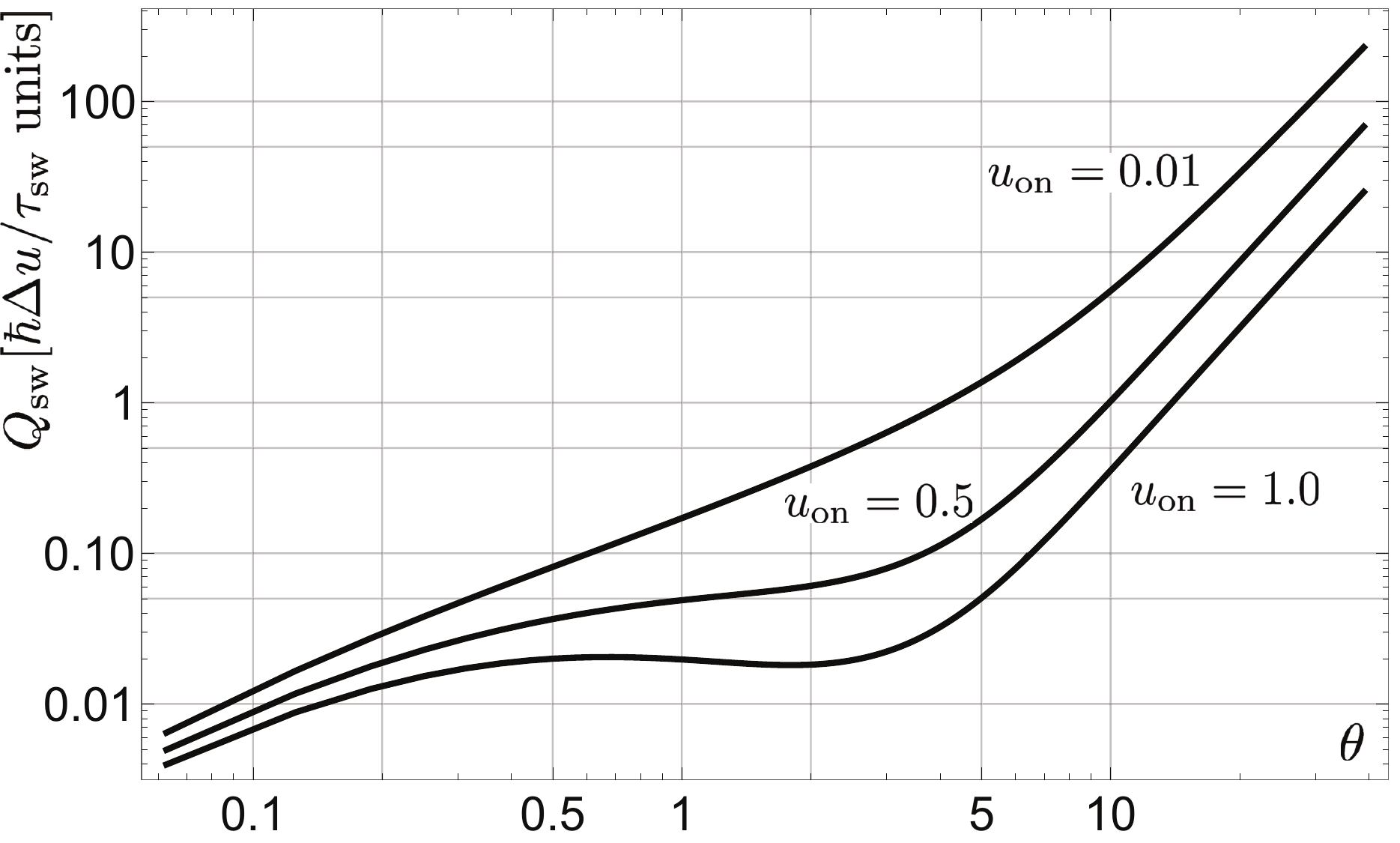}
  \caption{Switching heat in $\hbar \Delta u/\tau_\mathrm{sw}$ energy units (vertical axis, $\log$ scale)  versus the phase parameter $\theta$ (horizontal axis, $\log$ scale) for three different potential barrier heights $u_\mathrm{on}$.}\label{fig:q_vs_theta1}
\end{figure}

The dimensionless factor $F(\eta,u_\mathrm{on},\theta)$ strongly depends on the phase parameter $\theta$ as shown on Fig.\,\ref{fig:q_vs_theta1} for a specific value of $\eta = 0.25$ corresponding to $\tau_\varphi = 0.1$\,ps that is a typical electron-phonon inelastic scattering time in silicon at room temperature\,\cite{pop2010}. The phase parameter $\theta$ can be viewed as a ratio of the potential barrier length $L_{\mathrm g}$ and De Broglie wavelength $\lambda = h/mv_\mathrm{th}$ of a charge carrier moving with the thermal velocity $v_\mathrm{th} = \sqrt{2k_{\sr B}T/m}$,
\begin{equation}
      \theta = 2\pi\, \frac{L_g}{\lambda}.
\end{equation}

At $T=300$\,K, De Broglie wavelengths of electrons in silicon conduction band are $\lambda_t = 17.5$\,nm and $\lambda_\ell = 7.7$\,nm for transverse and longitudinal electron charge carrier masses $m_{\mathrm t} = 0.19m_{\mathrm e}$ and $m_\ell = 0.98m_{\mathrm e}$. Hence, for a large gate region with an extended potential barrier $\sim 50$\,nm the parameter $\theta$ is at most $\theta \approx 40$ for heavy electrons. At large $\theta \gg 1$ the factor $F$ scales as $\theta^\alpha$ with $\alpha \sim 3$ and can approach large values $F \sim 100$, see Fig.\,\ref{fig:q_vs_theta1}. Therefore, in order to minimize the switching heat one needs to reduce the gate region size as much as possible or, alternatively, increase the height of the potential barrier $u_\mathrm{on}$ in the open regime. This however conflicts with the requirement of a large current separation for the open and closed regimes of the transistor. The standby current $I_\mathrm{off}$ in the saturated regime of high bias voltages $V$ can be approximated as
\begin{equation}
      I_\mathrm{off} \approx \frac{e}{h}\Biggl(\int\limits_0^{U_\mathrm{off}} dE\, T(E,U_\mathrm{off}) n_{\sr F}(E) + \int\limits_{U_\mathrm{off}}^\infty dE\, n_{\sr F}(E)\Biggr),
\end{equation}
where the first term describes the underbarrier quantum tunneling and the second one stands for the thermally activated process. In the WKB approximation $T(E,U_\mathrm{off}) = \exp(-2L_{\mathrm g}\sqrt{2m(U_\mathrm{off}-E)}/\hbar)$ and therefore,
\begin{equation}
      I_\mathrm{off} \approx \frac{e}{\hbar} k_BT \Bigl( e^{-2\theta\sqrt{u_\mathrm{off}-1}}\bigl( i_\mathrm{on}-e^{-u_\mathrm{off}}\bigr)+e^{- u_\mathrm{off}}\Bigr),
\end{equation}
where $i_\mathrm{on} = \int_{u_\mathrm{on}}^\infty d\epsilon \, n_F(\epsilon)$. For a fixed $u_\mathrm{off} \gg 1$ the standby current is saturated at $\theta > \theta_\mathrm{cr}$ with the critical value
\begin{equation}
      \theta_\mathrm{cr} = \frac{u_\mathrm{off} + \ln(i_\mathrm{on} -e^{-u_\mathrm{off}})}{2\sqrt{u_\mathrm{off}-1}}.
\end{equation}
The saturated value of the on/off current ratio is
\begin{equation}
      \log\Bigl[\frac{I_\mathrm{on}}{I_\mathrm{off}}\Bigr] = u_\mathrm{off} \log(e) + \log(i_\mathrm{on}/2).
      \label{eq:ratio}
\end{equation}
For $u_\mathrm{on}$$\to$$0$ this ratio is expressed solely through the dimensionless potential barrier change $\Delta u$: $\log\bigl[I_\mathrm{on}/I_\mathrm{off}\bigr] \approx 0.43 \Delta u -0.46$.

The switching heat becomes important at high operation frequencies $\nu \sim (\Delta t)^{-1}$. Indeed, as far as $\Delta t$ decreases, the switching time $\tau_\mathrm{sw} \leq \Delta t$ has to decrease as well. On the contrary, the transport heat $\propto \Delta t$  decreases proportionally, see Eq.\,(\ref{eq:Qsd}). To quantitatively estimate the released switching heat, we take ten decades on/off currents difference, which, according to Eq.\,(\ref{eq:ratio}), requires $\Delta u \approx 24$. The minimal gate region where such a potential barrier height can be achieved is given by $\theta_\mathrm{cr} \approx 2.5$ corresponding to $L_{\mathrm g} = \lambda_{\mathrm t} \theta_\mathrm{cr}/(2\pi) \approx 7$\,nm. At this gate length, the main contribution to the switching heat is due to heavy electrons which gives
\begin{equation}
      Q_\mathrm{sw} = \frac{\hbar}{\tau_\mathrm{sw}}\, \Delta u\, F(\eta,u_\mathrm{on}\to 0, 2\pi\, \frac{L}{\lambda_\ell}) \approx 41\, \frac{\hbar}{\tau_\mathrm{sw}},
\end{equation}
per heavy electron conduction channel. Here we have taken $\eta=0.25$ ($\tau_\varphi = 0.1$\,ps). This heat has to be compared with the transport heat contribution given by Eq.\,(\ref{eq:Qsd}). In the following we choose $\Delta t = 3\tau_\mathrm{sw}$ implying that the device is open during $\tau_\mathrm{sw}$ time interval. Then
\begin{equation}
      Q_\mathrm{tr} = \frac{eV \tau_\mathrm{sw}}{2\pi\hbar} \int_0^\infty dE \,\bar{n}_E^{\rs} \approx 0.11\, k_{\sr B}T\,\frac{eV \tau_\mathrm{sw}}{\hbar}.
\end{equation}

\begin{figure}[ht]
  \centering
  \includegraphics[width=7.5cm]{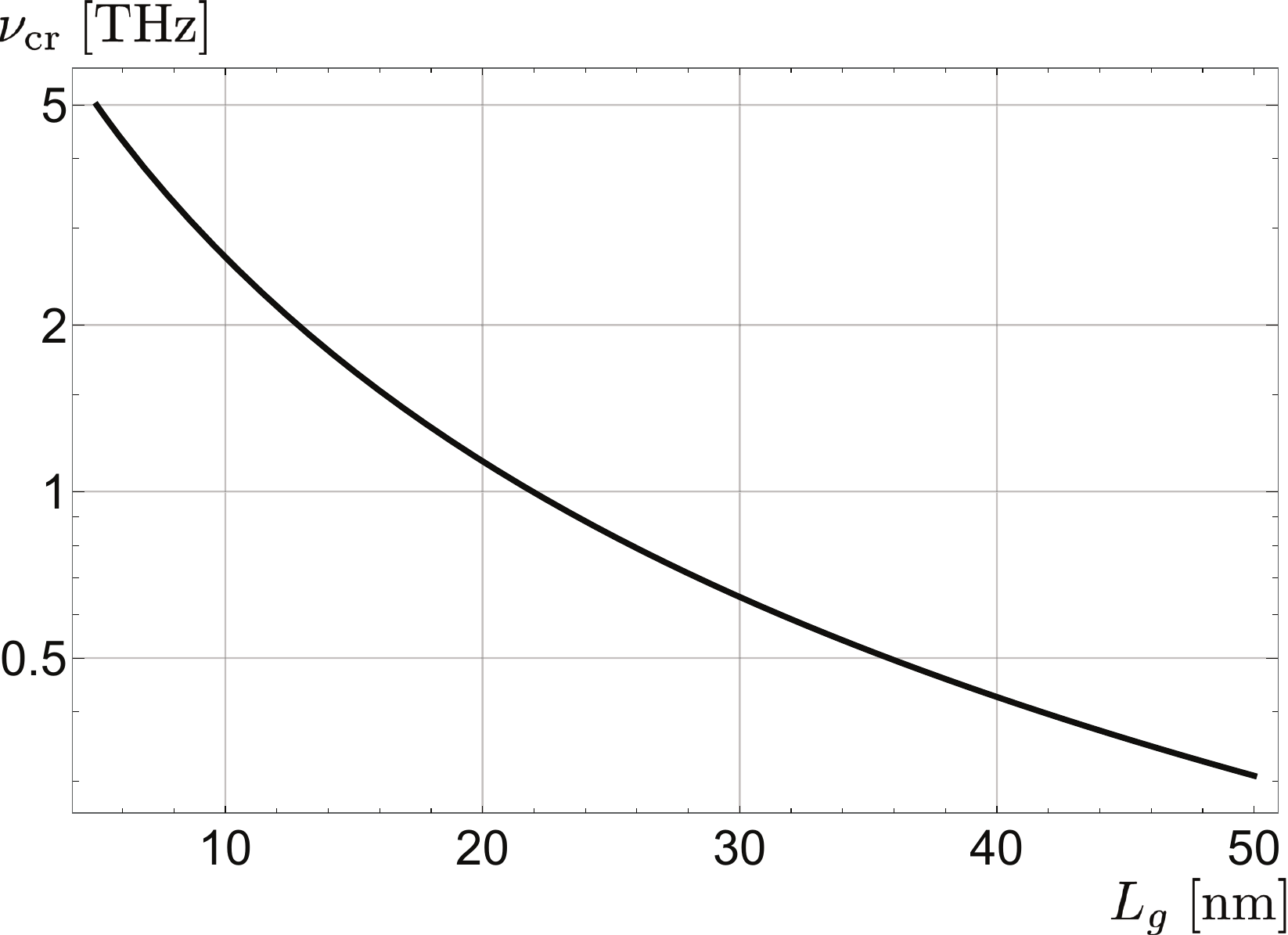}
  \caption{Critical operation frequency (vertical axis, $log$ scale) vs gate length for the silicon MOSFET device with $V=0.8$V and $\tau_\varphi=0.1$ps.}\label{fig:nucr}
\end{figure}

The characteristic operation frequency $\nu_\mathrm{cr}$ where the switching heat compares with the transport heat i.e. $Q_\mathrm{tr} = Q_\mathrm{sw}$ is given by
\begin{equation}
      \nu_\mathrm{cr} = \frac{k_{\sr B}T}{3\hbar}\sqrt{\frac{eV}{k_{\sr B}T}\, \frac{0.11}{\Delta u F}} \approx 3.77\, \mbox{THz},
\end{equation}
at typical supply voltage $0.8$\,V and room temperature conditions. At this frequency the switching heat is about $12k_{\sr B}T$ and $3.4k_{\sr B}T$ per heavy and light electron conduction channel correspondingly. In modern metal-oxide-semiconductor field-effect transistor (MOSFET) designs, the typical length of the gate region is larger than 7\,nm. Therefore, the device is not optimized with respect to the switching heat minimization. For such nonoptimized designs with  $L_{\mathrm g} \approx 50$\,nm, the switching heat is substantially larger, $146k_{\sr B}T$ and $14k_{\sr B}T$, per heavy and light electron conduction channels, respectively, at critical operation frequency $\nu_\mathrm{cr} \approx 300$GHz. This is far beyond the operation frequencies of modern digital devices. In the 10\,GHz frequency range with $\tau_\mathrm{sw} \simeq 10$\,ps typical for the current state-of-the-art digital MOSFETs, the switching heat gives only $\sim 0.1\%$ contribution to the net heat production. However, since the current trend in the semiconductor technology is to increase the operation frequency and lower the supply voltage, the AOC heating becomes more and more important and has to be taken into account in the prospective large integrated circuit designs. The critical operation frequency as a function of the gate channel length for the silicon MOSFET device with $V=0.8$\,V and $\tau_\varphi=0.1$\,ps is shown in Fig.\,\ref{fig:nucr}. The above analysis holds for ballistic devices with the relatively short $L_{\mathrm g} \sim 100$\,nm. At larger $L_{\mathrm g}$, the electron transport is diffusive and electrons lose their coherence during the propagation inside the channel region.

In conclusion, building on the scattering matrix approach, we derived a general expression for the released heat in the voltage biased mesoscopic conductor with the time dependent scattering matrix. The released heat is a sum of two different contributions, the transport heat, associated with the passage of electrons driven by the bias voltage, see Eq. (\ref{eq:Qtr}), and the switching heat originating from the change in the scattering potential, see Eq.\,(\ref{eq:Qsw}). The latter has an intrinsic quantum origin and is a manifestation of the Anderson catastrophe, i.e. a result of the adjustment of the many-particle electron wave function to the new scattering potential. The change in the potential generates the electron-hole excitations in the system, which results in the additional heat generation. For a finite coherence time where the scattering matrix of the system changes slowly on the coherence time temporal scale, the switching heat is given by the square of the first-order derivatives of the scattering coefficients, see Eq.\,(\ref{eq:Qsw3}). Importantly, the switching heat is always present even in a non-biased system, $V=0$, and, therefore, this heat has to be taken into account as far as the supply voltage and switching times are further reduced.

The work of A.V.L. was supported by Bel Huawei Technology, LLC. under the research Contract No. YBN2019075147 and by the RFBR Grant No. 18-02-00642A.  The work of V.M.V. at Argonne was supported by
the U.S. Department of Energy, Office of Science, Basic
Energy Sciences, Materials Sciences and Engineering
Division.

\appendix

\begin{widetext}
\section{Non-equilibrium electron occupation numbers} \label{A1}

We assume that the scattering amplitudes change slowly on a time scale of the ballistic time flight of the electron through the scattering region.
Then in the WKB approximation the non-stationary scattering states for the electrons with the kinetic energy $E$ coming from the source and drain reservoirs are given by,
\begin{eqnarray}
      &&\varphi_E^{\rs}(x,t) = \theta(-x)\Bigl( e^{ik_E^{\rs}x} +r_E(\tau_{\s +}) e^{-ik_E^{\rs}x} \Bigr)e^{-iEt/\hbar}
      + \theta(x) \sqrt{\frac{k_E^{\rs}}{k_E^{\rd}}} t_E(\tau_{\s -}) e^{ik_E^{\rd} x-iEt/\hbar},
      \label{eq:SSA}
      \\
      &&\varphi_E^{\rd}(x,t) = \theta(-x) \sqrt{\frac{k_E^{\rd}}{k_E^{\rs}}} t_E(\tau_{\s +}) e^{-ik_E^{\rs}x-iEt/\hbar}
      +\theta(x) \Bigl(e^{-ik_E^{\rd}x} + \bar{r}_E(\tau_{\s -}) e^{ik_E^{\rd} x} \Bigr) e^{-iEt/\hbar},
      \label{eq:SSB}
\end{eqnarray}
\end{widetext}
where $\theta(\pm x)$ are Heaviside functions, $k_E^{\ri} = \sqrt{2m\bigl[E-E_\mathrm{c}^{\ri}\bigr]}/\hbar$ is the electron wavevector in the terminal $\mathrm{i}\in\{\mathrm{s,d}\}$ with the bottom of the conduction band $E_\mathrm{c}^{\ri}$. The time-dependent scattering amplitudes $t_E(\tau_{\s \pm})$ with $\tau_{\s +} = \tau + x/v_E^{\rs}$  and $\tau_{\s -} = \tau - x/v_E^{\rd}$ describe the ballistically time-retarded transmission of an electron at energy $E$ between the source and the drain terminals, whereas $r_E(\tau_{\s \pm})$ and $\bar{r}_E(\tau_{\s \pm})$ describe the back-reflection scattering to the source and drain terminals correspondingly, $v_E^{\ri} = \hbar k_E^{\ri}/m$ is the electron velocity in the terminal $\mathrm{i}\in\{\mathrm{s,d}\}$. The stationary scattering states for the 'OFF' regime are given by Eqs.(\ref{eq:SSA}) and (\ref{eq:SSB}) with all time-dependent scattering amplitudes replaced by the stationary scattering amplitudes for the `OFF' regime: $t_E(\tau_{\s \pm}) \to t_E^{\s (0)}$, $r_E(\tau_{\s +}) \to r_E^{\s (0)}$ and $\bar{r}_E(\tau_{\s -}) \to \bar{r}_E^{\s (0)}$.

Substituting Eqs.(\ref{eq:SSA}) and (\ref{eq:SSB}) into Eq.(\ref{eq:ab_transform}) and keeping only slow oscillating exponential terms of the form $\exp\bigl[\pm i(k_{E^\prime}^{\ri}-k_E^{\ri})x\bigr] \approx \exp\bigl[\pm i (E^\prime -E)x/(\hbar v_{E^\prime}^{\ri})\bigr]$ one gets the non-stationary electronic annihilation operators,
\begin{widetext}
\begin{eqnarray}
      &&\hat{a}_E(t) = \sum_{E^\prime} v_{E^\prime}^{\rs} \hat{\bar a}_{E^\prime} \Biggl( \int\limits_t^\infty d\tau\, e^{-i(E^\prime -E)\tau/\hbar} + \int\limits_{-\infty}^t d\tau \Bigl( r_{E^\prime}(\tau)r_E^{*\s(0)} + \sqrt{\frac{v_{E^\prime}^{\rd} v_E^{\rs}}{v_{E^\prime}^{\rs} v_E^{\rd}}} \,t_{E^\prime}(\tau) t_E^{*\s(0)}\Bigr) e^{-i(E^\prime-E)\tau/\hbar} \Biggr)
      \label{eq:a}\\
      &&\qquad\>\> + \sum_{E^\prime} \sqrt{v_{E^\prime}^{\rs} v_{E^\prime}^{\rd}} \,\hat{\bar b}_{E^\prime} \int\limits_{-\infty}^t d\tau \Bigl( t_{E^\prime}(\tau)r_E^{*\s(0)} + \sqrt{\frac{v_{E^\prime}^{\rd} v_E^{\rs}}{v_{E^\prime}^{\rs} v_E^{\rd}}} \,\bar{r}_{E^\prime}(\tau) t_E^{*\s(0)}\Bigr) e^{-i(E^\prime - E)\tau/\hbar},
      \nonumber
      \\
      &&\hat{b}_E(t) = \sum_{E^\prime} \sqrt{v_{E^\prime}^{\rs} v_{E^\prime}^{\rd}} \,\hat{\bar a}_{E^\prime} \int\limits_{-\infty}^t d\tau \Bigl( \sqrt{\frac{v_{E^\prime}^{\rs} v_E^{\rd}}{v_{E^\prime}^{\rd} v_E^{\rs}}} \,r_{E^\prime}(\tau) t_E^{*\s(0)} + t_{E^\prime}(\tau) \bar{r}_E^{*\s(0)}\Bigr) e^{-i(E^\prime - E)\tau/\hbar}
      \label{eq:b}\\
      &&\qquad\>\> + \sum_{E^\prime} v_{E^\prime}^{\rd}\, \hat{\bar b}_{E^\prime} \Biggl( \int\limits_t^\infty d\tau e^{-i(E^\prime - E)\tau/\hbar} + \int\limits_{-\infty}^t d\tau\, \Bigl(\sqrt{\frac{v_{E^\prime}^{\rs} v_E^{\rd}}{v_{E^\prime}^{\rd} v_E^{\rs}}} \,t_{E^\prime}(\tau) t_E^{*\s(0)} + \bar{r}_{E^\prime}(\tau) \bar{r}_E^{*\s(0)}\Bigr) e^{-i(E^\prime-E)\tau/\hbar}\Biggr).
      \nonumber
\end{eqnarray}
Next, we note that if one replaces all time-dependent scattering amplitudes in Eqs.(\ref{eq:a}) and (\ref{eq:b}) by the corresponding stationary amplitudes in the 'OFF' regime, then due to orthogonality of the scattering states $\bar\varphi^{\rs}_E$ and $\bar\varphi^{\rd}_E$ one has to get: $\hat{a}_E(t) = \hat{\bar a}_E$ and $\hat{b}_E(t) = \hat{\bar b}_E$. Therefore one can rewrite the non-stationary annihilation operators in the form,
\begin{eqnarray}
      &&\hat{a}_E(t) = \sum_{E^\prime} v_{E^\prime}^{\rs} \hat{\bar a}_{E^\prime} \Bigl( 2\pi\hbar\delta(E^\prime\!-\!E)
      +\!\!\! \int\limits_{-\infty}^t\!\!\! d\tau F_{E,E^\prime}^{\rs}(\tau) e^{i(E-E^\prime)\tau/\hbar}\Bigr) +\! \sum_{E^\prime}\! \sqrt{v_{E^\prime}^{\rs} v_{E^\prime}^{\rd}} \hat{\bar b}_{E^\prime} \!\!\!\int\limits_{-\infty}^t\!\!\! d\tau G_{E,E^\prime}^{\rs}(\tau) e^{i(E-E^\prime)\tau/\hbar},
      \\
      &&\hat{b}_E(t) = \sum_{E^\prime} \sqrt{v_{E^\prime}^{\rs} v_{E^\prime}^{\rd}} \hat{\bar a}_{E^\prime} \!\!\!\int\limits_{-\infty}^t\!\!\! d\tau G_{E,E^\prime}^{\rd}(\tau) e^{i(E-E^\prime)\tau/\hbar} + \sum_{E^\prime}v_{E^\prime}^{\rd} \hat{\bar b}_{E^\prime} \Bigl( 2\pi\hbar\delta(E^\prime\!-\!E)+\!\!\!\int\limits_{-\infty}^t\!\!\! d\tau F_{E,E^\prime}^{\rd}(\tau) e^{i(E-E^\prime)\tau/\hbar} \Bigr),
\end{eqnarray}
\end{widetext}
where we have defined,
\begin{equation}
      F_{E,E^\prime}^{\rs}\!(\tau) \!=\! \delta r_{E^\prime}(\tau)r_E^{*\s(0)} \!\!+\!  \sqrt{\frac{v_{E^\prime}^{\rd} v_E^{\rs}}{v_{E^\prime}^{\rs} v_E^{\rd}}} \,\delta t_{E^\prime}(\tau) t_E^{*\s(0)},
\end{equation}
\begin{equation}
      F_{E,E^\prime}^{\rd}\!(\tau) \!=\! \sqrt{\frac{v_{E^\prime}^{\rs} v_E^{\rd}}{v_{E^\prime}^{\rd} v_E^{\rs}}} \,\delta t_{E^\prime}(\tau) t_E^{*\s(0)} \!\!+\! \delta \bar{r}_{E^\prime}(\tau) \bar{r}_E^{*\s(0)},
\end{equation}
\begin{equation}
      G_{E,E^\prime}^{\rs}\!(\tau) \!=\! \delta t_{E^\prime}(\tau)r_E^{*\s(0)} \!\!+\! \sqrt{\frac{v_{E^\prime}^{\rd} v_E^{\rs}}{v_{E^\prime}^{\rs} v_E^{\rd}}} \,\delta\bar{r}_{E^\prime}(\tau) t_E^{*\s(0)},
\end{equation}
\begin{equation}
      G_{E,E^\prime}^{\rd}\!(\tau) \!=\! \sqrt{\frac{v_{E^\prime}^{\rs} v_E^{\rd}}{v_{E^\prime}^{\rd} v_E^{\rs}}} \delta r_{E^\prime}(\tau) t_E^{*\s(0)} \!\!+\! \delta t_{E^\prime}(\tau) \bar{r}_E^{*\s(0)},
\end{equation}
with $\delta t_E(\tau) \equiv t_E(\tau) - t_E^{\s (0)}$, $\delta r_E(\tau) = r_E(\tau) - r_E^{\s(0)}$ and $\bar{r}_E(\tau) = \bar{r}_E(\tau)-\bar{r}_E^{\s(0)}$ are variations of the scattering amplitudes which are nonvanishing only during `ON' regime of the device.

The resulting electron energy distribution can be found from the single-electron density matrices: $\rho_{EE^\prime}^{\rs}(t) \equiv \langle \hat{a}_E^\dagger(t) \hat{a}_{E^\prime}(t)\rangle$ and $\rho_{EE^\prime}^{\rd}(t) \equiv \langle \hat{b}_E^\dagger(t) \hat{b}_{E^\prime}(t)\rangle$. In particular, we are interested in the electron occupation numbers $N_E^{\ri}(t) \equiv \lim\limits_{E^\prime \to E} \rho_{E,E^\prime}^{\ri}(t)$, $\mathrm{i}\in\{\mathrm{s,d}\}$. The variations of the occupation numbers, $\delta n_E^{\ri} \equiv (N_E^{\ri}/L - \bar{n}_E^{\ri})$ are then given by
\begin{widetext}
\begin{equation}
      \delta n_E^{\rs}= \bar{n}_E^{\rs} \!\!\!\int\!\! d\tau \Bigr\{F_{E,E}^{\rs}(\tau) +F_{E,E}^{*\rs}(\tau)\Bigr\} + \!\!\int\!\! d\tau d\tau^\prime \sum_{E^\prime} e^{i(E^\prime-E)(\tau-\tau^\prime)/\hbar}
      \Bigl\{F_{E,E^\prime}^{*\rs}(\tau) F_{E,E^\prime}^{\rs}(\tau^\prime) \bar{n}_{E^\prime}^{\rs} + G_{E,E^\prime}^{*\rs}(\tau) G_{E,E^\prime}^{\rs}(\tau^\prime) \bar{n}_{E^\prime}^{\rd}\!\Bigr\},
      \label{eq:NA}
\end{equation}
\end{widetext}
and similarly for the drain terminal. In the following we assume that the switching time $\tau_\mathrm{sw}$ during which the scattering amplitudes change from the 'OFF' to the 'ON' regime and the back is large enough so that the associated Heisenberg energy scale $\Delta E = \hbar/\tau_\mathrm{sw}$ is less than the characteristic scale $\Gamma$ of energy dependence of the scattering amplitudes. Then one can assume $E=E^\prime$ under time integrals in Eq.(\ref{eq:NA}) and get,
\begin{eqnarray}
     &&\delta n^{\rs}_E = \frac{v_E^{\rs}}{L}\int \frac{dE^\prime}{2\pi\hbar} \int d\tau d\tau^\prime\, e^{i(E^\prime-E)(\tau-\tau^\prime)/\hbar}
     \label{eq:dnA}\\
     &&\>\>\> \times\Bigl\{\bar{n}^{\rs}_{E^\prime} \bigl[ r_{E^\prime}^*(\tau) r_{E^\prime}(\tau^\prime) + t_{E^\prime}^{*}(\tau)t_{E^\prime}(\tau^\prime)-1\bigr]
     \nonumber\\
     &&\quad-\bar{n}_{E^\prime}^{\rs} f^{*\rs}_{E^\prime}(\tau) f^{\rs}_{E^\prime}(\tau^\prime) + \bar{n}_{E^\prime}^{\rd} f^{*\rd}_{E^\prime}(\tau) f^{\rd}_{E^\prime}(\tau^\prime)\Bigr\},
     \nonumber
\end{eqnarray}
\begin{eqnarray}
     &&\delta n^{\rd}_E = \frac{v_E^{\rd}}{L}\int \frac{dE^\prime}{2\pi\hbar} \int d\tau d\tau^\prime\, e^{i(E^\prime-E)(\tau-\tau^\prime)/\hbar}
     \label{eq:dnB}\\
     &&\>\>\>\times\Bigl\{\bar{n}^{\rd}_{E^\prime} \bigl[ \bar{r}_{E^\prime}^*(\tau) \bar{r}_{E^\prime}(\tau^\prime) + t_{E^\prime}^{*}(\tau)t_{E^\prime}(\tau^\prime)-1\bigr]
     \nonumber\\
     &&\quad+\bar{n}_{E^\prime}^{\rs} f^{*\rs}_{E^\prime}(\tau) f^{\rs}_{E^\prime}(\tau^\prime) - \bar{n}_{E^\prime}^{\rd} f^{*\rd}_{E^\prime}(\tau) f^{\rd}_{E^\prime}(\tau^\prime)\Bigr\},
     \nonumber
\end{eqnarray}
with
\begin{eqnarray}
      &&f^{\rs}_E(\tau) \equiv t_E(\tau)\bar{r}_E^{*\s(0)} +r_E(\tau)t_E^{*\s(0)}
      \\
      &&f^{\rd}_E(\tau) \equiv t_E(\tau) r_E^{*\s(0)} + \bar{r}_E(\tau)t_E^{*\s(0)}.
\end{eqnarray}
These functions vanish at $|t|\gg \Delta t/2$ due to unitarity constraint $t_E^{\s(0)}\bar{r}_E^{*\s(0)} +r_E^{\s(0)}t_E^{*\s(0)} = 0$, whereas at coincident time instants $|f^{\rs}_E(\tau)|^2 = |f^{\rd}_E(\tau)|^2$.

\end{document}